\documentclass[superscriptaddress, letterpaper,twocolumn,prl,showpacs,reprint]{revtex4-1}

\usepackage{graphicx,epsfig}
\usepackage{dcolumn}
\usepackage{bm}
\usepackage{textcomp}

\begin{document}

\title{Strain Control of Fermiology and Many-Body Interactions in Two-Dimensional Ruthenates}

\author{B. Burganov}
\affiliation{Laboratory of Atomic and Solid State Physics, Department of Physics, Cornell University, Ithaca, New York 14853, USA}

\author{C. Adamo}
\affiliation{Department of Materials Science and Engineering, Cornell University, Ithaca, New York 14853, USA}
\affiliation{Department of Applied Physics, Stanford University, Stanford, CA 94305, USA} 

\author{A. Mulder}
\affiliation{School of Applied and Engineering Physics, Cornell University, Ithaca, New York 14853, USA}

\author{M. Uchida}
\affiliation{Laboratory of Atomic and Solid State Physics, Department of Physics, Cornell University, Ithaca, New York 14853, USA}

\author{P. D. C. King}
\affiliation{Laboratory of Atomic and Solid State Physics, Department of Physics, Cornell University, Ithaca, New York 14853, USA}
\affiliation{School of Physics and Astronomy, University of St Andrews, St Andrews, Fife KY16 9SS, UK}
\affiliation{Kavli Institute at Cornell for Nanoscale Science, Ithaca, New York 14853, USA}

\author{J. W. Harter}
\affiliation{Laboratory of Atomic and Solid State Physics, Department of Physics, Cornell University, Ithaca, New York 14853, USA}

\author{D. E. Shai}
\affiliation{Laboratory of Atomic and Solid State Physics, Department of Physics, Cornell University, Ithaca, New York 14853, USA}

\author{A. S. Gibbs}
\affiliation{Max Planck Institute for Solid State Research, 70569 Stuttgart, Germany}

\author{A. P. Mackenzie}
\affiliation{School of Physics and Astronomy, University of St Andrews, St Andrews, Fife KY16 9SS, UK}
\affiliation{Max Planck Institute for Chemical Physics of Solids, D-01187 Dresden, Germany}

\author{R. Uecker}
\affiliation{Leibniz Institute for Crystal Growth, D-12489 Berlin, Germany}

\author{M. Bruetzam}
\affiliation{Leibniz Institute for Crystal Growth, D-12489 Berlin, Germany}

\author{M. R. Beasley}
\affiliation{Department of Applied Physics, Stanford University, Stanford, CA 94305, USA} 

\author{C. J. Fennie}
\affiliation{School of Applied and Engineering Physics, Cornell University, Ithaca, New York 14853, USA}

\author{D. G. Schlom}
\affiliation{Department of Materials Science and Engineering, Cornell University, Ithaca, New York 14853, USA}
\affiliation{Kavli Institute at Cornell for Nanoscale Science, Ithaca, New York 14853, USA}

\author{K. M. Shen}
\email[Author to whom correspondence should be addressed: ]{kmshen@cornell.edu}
\affiliation{Laboratory of Atomic and Solid State Physics, Department of Physics, Cornell University, Ithaca, New York 14853, USA}
\affiliation{Kavli Institute at Cornell for Nanoscale Science, Ithaca, New York 14853, USA}


\begin{abstract}
Here we demonstrate how the Fermi surface topology and quantum many-body interactions can be manipulated via epitaxial strain in the spin-triplet superconductor Sr$_2$RuO$_4$ and its isoelectronic counterpart Ba$_2$RuO$_4$ using oxide molecular beam epitaxy (MBE), \emph{in situ} angle-resolved photoemission spectroscopy (ARPES), and transport measurements. Near the topological transition of the $\gamma$ Fermi surface sheet, we observe clear signatures of critical fluctuations, while the quasiparticle mass enhancement is found to increase rapidly and monotonically with increasing Ru-O bond distance. Our work demonstrates the possibilities for using epitaxial strain as a disorder-free means of manipulating emergent properties, many-body interactions, and potentially the superconductivity in correlated materials.
\end{abstract}

\pacs{74.70.Pq, 74.25.Jb, 79.60.Dp}

\maketitle

\begin{figure*}[tbp]
\begin{center}
\includegraphics[width=7in]{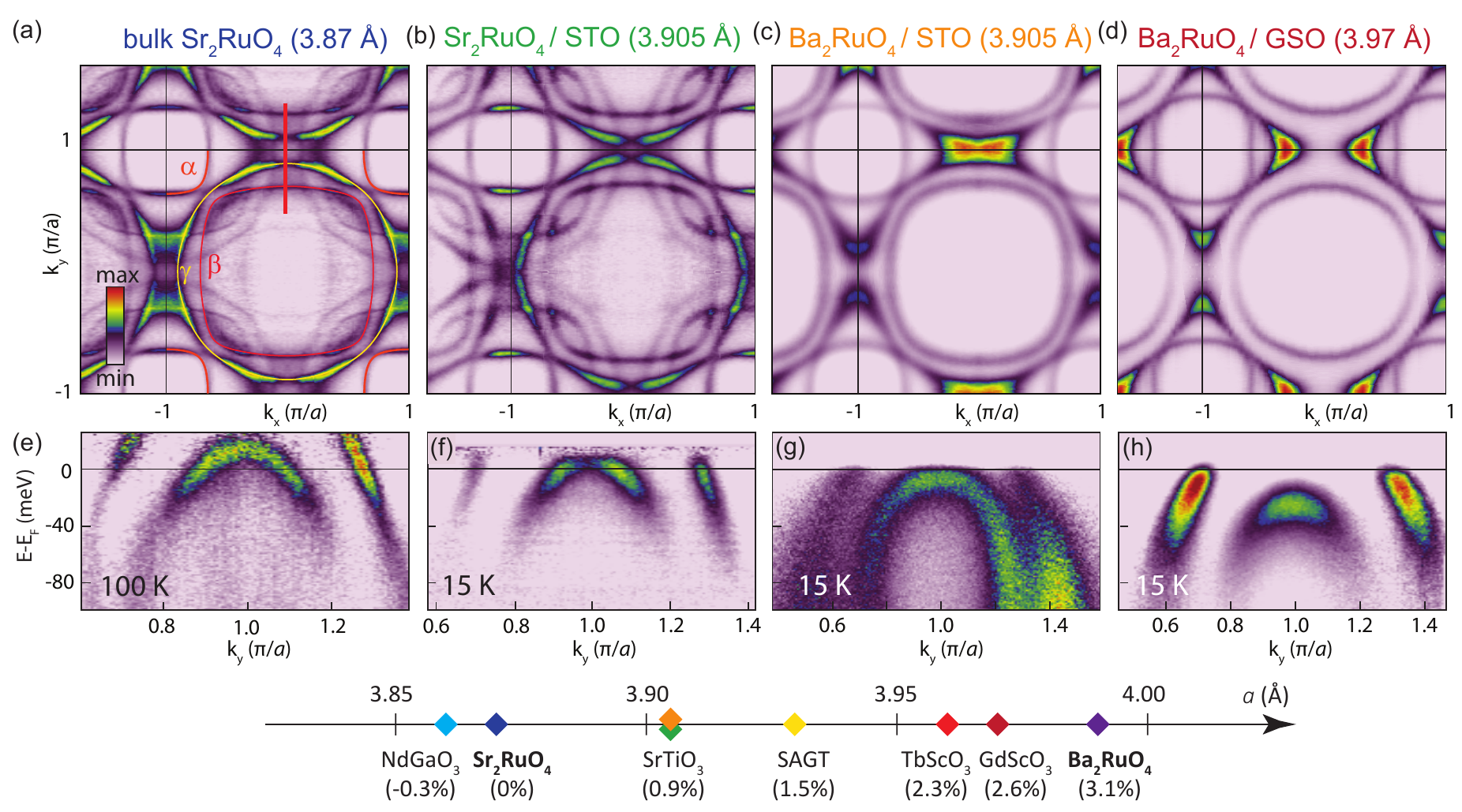}
\caption{
(a)-(d) Fermi surface maps and (e)-(h) spectral weight along the $(0,k_y)$ direction (thick red line in (a)) for select strain states of Sr$_2$RuO$_4$ and Ba$_2$RuO$_4$. 
The data in (e) was measured at an elevated temperature (T = 100 K) to thermally populate the states above the Fermi level; the rest of the data in the paper was taken at 15 K. To show the dispersion near the vHs above $\mathrm{E_F}$, the spectral weight was divided by the Fermi function in (e) and (f). 
Substrate number line shows the room temperature lattice constants and strain values relative to  bulk Sr$_2$RuO$_4$.
}\label{fig:fig1}
\end{center}
\end{figure*}

Pressure plays a key role in modifying the properties of materials with strong electronic correlations, for instance, enhancing the transition temperature of the cuprate superconductors or driving quantum phase transitions in heavy fermion systems. Unfortunately, leading techniques for investigating the electronic structure, such as ARPES and STM, are incompatible with typical high pressure / strain apparatus. The epitaxial growth of thin films on deliberately lattice mismatched substrates provides a clean and accessible analogue to external pressure and has been used to dramatically alter the electronic phases of many complex oxides \cite{locquet1998, choi2004, leejh2010, YooHK2015}. 
In the family of ruthenium oxides, the strong structure-property relationship leads to a wide variety of ground states including unconventional superconductivity\cite{Maeno1994}, metamagnetism and electronic liquid crystalline states \cite{Grigera2001, Borzi2007, Lester2015}, ferromagnetism, antiferromagnetism and spin-glass behavior \cite{Longo1968, Nakatsuji2000,Carlo2012}, without changing the formal oxidation state of the Ru ion.
Among them, Sr$_{2}$RuO$_{4}$ is an ideal candidate to explore the effects of biaxial strain and chemical pressure, since the extreme sensitivity of its superconducting ground state to disorder\cite{Mackenzie1998a} precludes enhancement of $\mathrm{T_{c}}$ through chemical substitution.
The possibly chiral nature of the superconducting state has given rise to proposals utilizing Sr$_{2}$RuO$_{4}$ as a platform for realizing Majorana fermions, exotic Josephson junctions, and non-Abelian topological quantum computation \cite{Ueno2013, Brydon2010}. Hydrostatic pressure was shown to suppress both the $\mathrm{T_{c}}$ \cite{Shirakawa1997} and quasiparticle enhancements \cite{Forsythe2002}, but recent experiments applying a uniaxial strain of $0.2\%$ demonstrated a strong nonlinear enhancement of $\mathrm{T_{c}}$ \cite{Hicks2014}. 
Obtaining uniaxial strains of greater than 0.5\% is a challenge in rather brittle metal oxides, but biaxial strains of 2-3\% are readily achievable in epitaxial thin films grown on deliberately lattice mismatched substrates. Here we demonstrate epitaxial strain engineering as a disorder-free means to dramatically manipulate the electronic structure of Sr$_2$RuO$_4$ and its sister compound, Ba$_2$RuO$_4$, through a combination of reactive oxide molecular beam epitaxy (MBE) growth and \emph{in situ} ARPES. We are able to observe a topological transition in the $\gamma$ Fermi surface (FS) sheet (i.e., a Lifshitz transition) through the selection of appropriate substrates. In addition, we observe signatures of quantum criticality in both ARPES and electrical transport near the Lifshitz transition, as well as a surprisingly large enhancement of the quantum many-body interactions with increasing in-plane lattice constant. 

\begin{figure}
\includegraphics[width=3.375in]{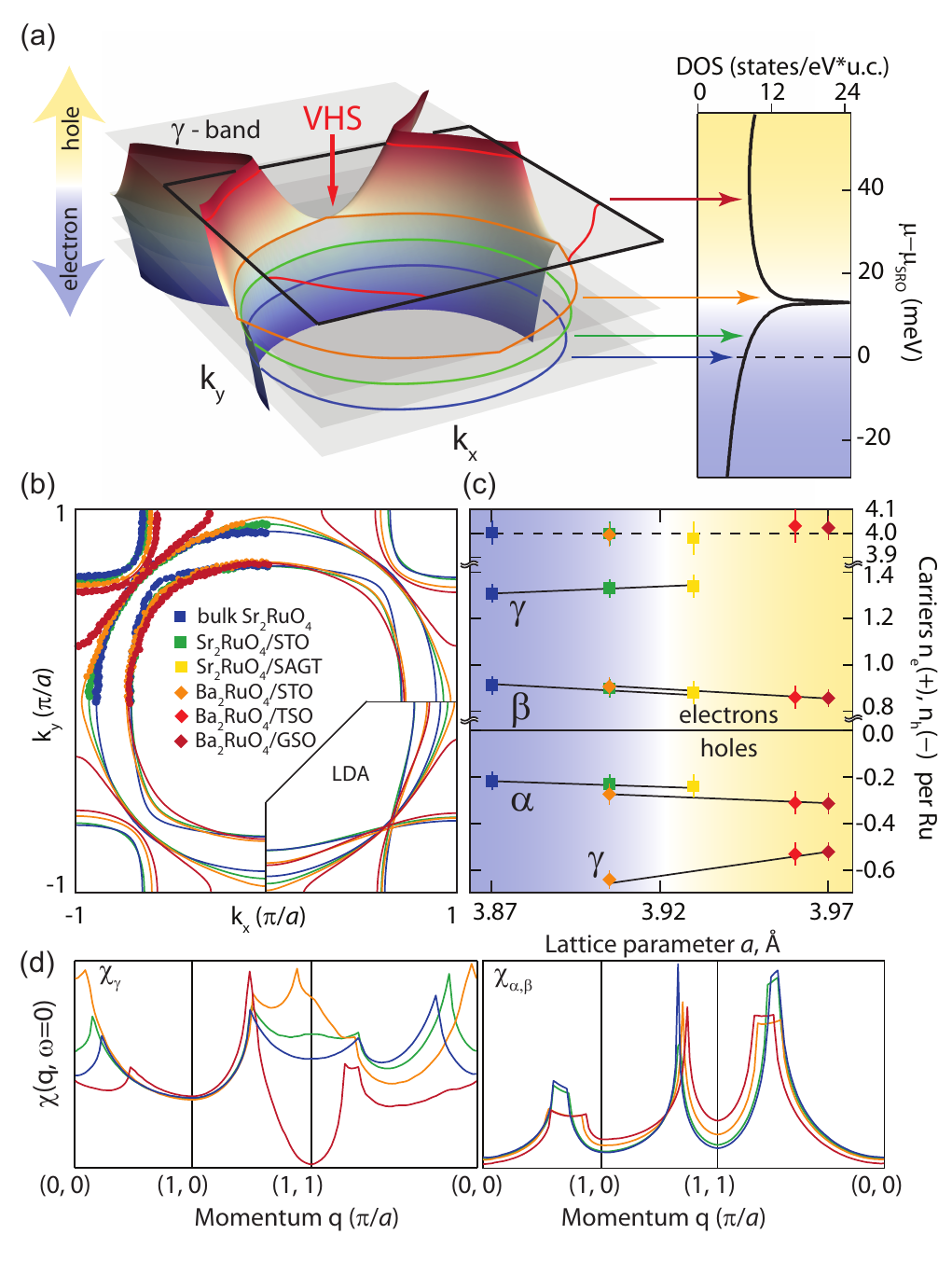}
\caption{
(a) Schematic showing the evolution of the $\gamma$ Fermi surface and density of states at $\mathrm{E_F}$ as a result of strain and negative chemical pressure by A-site substitution. (b) Tight-binding parametrization of ARPES Fermi surfaces and LDA Fermi surfaces. (c) Luttinger volume of experimental Fermi pockets as a function of the in-plane lattice parameter. The total number of electrons adds up to n=4.00$\pm0.05$ showing negligible overall doping. (d) Lindhard susceptibility calculated for the two-dimensional $\gamma$ and one-dimensional $\alpha$/$\beta$ pockets from the experimental Fermi surfaces.  
\label{fig:fig2}}
\end{figure}

Thin films of Sr$_2$RuO$_4$ and Ba$_2$RuO$_4$ were synthesized by reactive oxide MBE and the in-plane lattice constant (i.e., the Ru-O-Ru bond distance) can be increased from 3.87~\AA\ to 3.97~\AA\ ($\Delta a / a = 2.6\%$) through the selection of appropriate substrates. Sr$_2$RuO$_4$ films were found to relax immediately at lattice constants larger than 3.91 \AA, thus necessitating the substitution of Ba for Sr as the A-site cation to achieve even larger in-plane lattice constants. In bulk, Ba$_2$RuO$_4$ crystallizes in a hexagonal polymorph, and the K$_2$NiF$_4$ structure is metastable and has only been synthesized in polycrystalline form above 6 GPa \cite{Kafalas1972}. Epitaxial stabilization has, however, been employed to realize thin films of tetragonal Ba$_2$RuO$_4$ \cite{Jia1999} which we show are isostructural and isoelectronic to Sr$_2$RuO$_4$. 

The electronic structure of bulk Sr$_2$RuO$_4$ is highly two-dimensional and comprised of four electrons in the Ru $4d$ $t_{2g}$ orbitals, which form the quasi-1D $\alpha$ and $\beta$ FS sheets (primarily of $d_{xz}$ and $d_{yz}$ character), and the quasi-2D $\gamma$ sheet (primarily $d_{xy}$).  In Figure 1, we show a series of ARPES FS maps as a function of increasing in-plane lattice constant on a bulk single crystal of Sr$_2$RuO$_4$ cleaved at elevated temperature ($a =3.869$ \AA), Sr$_2$RuO$_4$ grown on SrTiO$_3$ (STO; $a =3.905$ \AA), Ba$_2$RuO$_4$ grown on SrTiO$_3$ ($a = 3.905$ \AA), and Ba$_2$RuO$_4$ grown on GdScO$_3$ (GSO; $a = 3.968$ \AA). The data from the single crystal of Sr$_2$RuO$_4$ (Fig. 1a) shows all three bulk FS sheets, as well as a $\sqrt{2} \times \sqrt{2}$ surface reconstruction, which generates additional sets of folded surface-derived bands\cite{Matzdorf2000,Damascelli2001}. The $\sqrt{2} \times \sqrt{2}$ surface reconstruction is still apparent in Fig. 1b, indicating that the reconstruction is also present on the natively grown surface. One of the unique hallmarks of Sr$_2$RuO$_4$ is the presence of a saddle point at $(\pi,0)$ and $(0, \pi)$, which gives rise to a van Hove singularity (vHs) only 14 meV above the Fermi energy (E$_{F}$, Fig. 1e). When this vHs passes through E$_{F}$, the $\gamma$ sheet undergoes a topological transition from electron-like to hole-like. For the thin film of Sr$_2$RuO$_4$ / STO (Fig. 1b), the $\gamma$ FS sheet is noticeably enlarged versus bulk and the vHs is pushed down to 9 meV above E$_{F}$. 

For Ba$_2$RuO$_4$ on SrTiO$_3$ (Fig. 1c), the $\gamma$ FS is almost precisely at the topological transition between electron and hole-like, and the vHs is nearly at E$_{F}$ (4 meV below, Fig. 1g). Although the samples shown in Fig. 1b and 1c are both grown on SrTiO$_{3}$, Ba$_2$RuO$_4$ / SrTiO$_{3}$ is much closer to the topological transition primarily due to the reduced second nearest neighbor hopping ($t_{4}$/$t_{1}$) which changes the shape of the $\gamma$ FS and lowers the vHs \cite{Supplemental}. For Ba$_2$RuO$_4$ grown on GdScO$_3$, the vHs is now well below E$_{F}$ (25 meV below, Fig. 1h), and the $\gamma$ FS clearly forms a hole-like sheet centered around $(\pi,\pi)$. The surface reconstruction is absent in Ba$_2$RuO$_4$ films, likely due to the larger Ba cation radius (Ba$^{2+}$ : 1.47 \AA\ vs. Sr$^{2+}$ : 1.31 \AA \cite{Shannon1976}) which should impede the freezing of the $\Sigma_3$ phonon mode on the surface. It is also notable that the $\beta$ FS sheet becomes noticeably less 1D, due to the increased transverse hopping between $d_{xz/yz}$ orbitals ($t_{3}/t_{2}$).

\begin{figure}
\begin{center}
\includegraphics[width=3.375in]{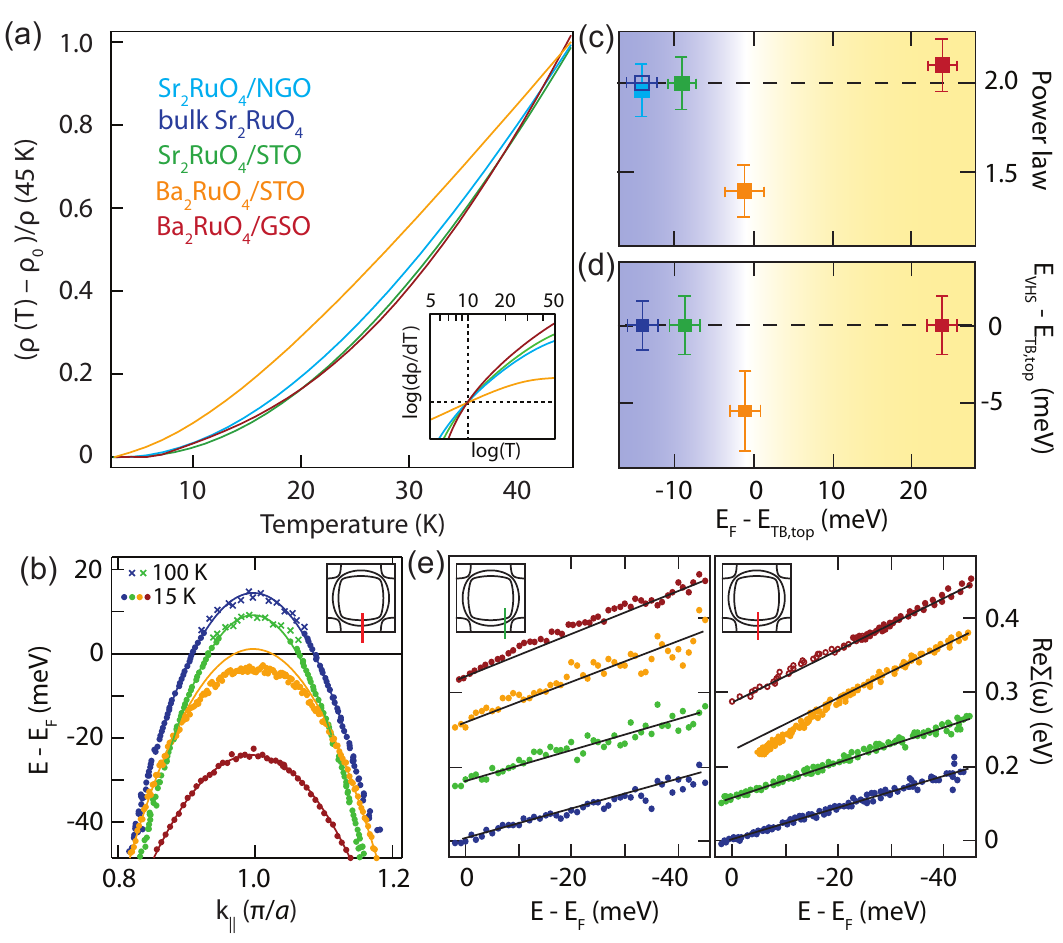}
\caption{ Normalized resistivity fitted to $\rho \propto T^{n}$ (values of $n$ shown in (c) with the open symbol from \cite{Hussey1998}). The inset shows $\log(d\rho/dT) \approx (n-1) \log(T)$ with an offset. (b) Dispersion of the $\gamma$-band along $(0,0)-(0,\pi$), and (d) the deviation of $\mathrm{E_{VHS}}$ from the TB model. (e) Re$\Sigma (\omega)$ (offset at $\mathrm{E=E_F}$ is included for clarity). $\mathrm{Re}\Sigma (\omega)\propto \omega$ implies quadratic energy dependence of the quasiparticle scattering rate $\Gamma (\omega)\propto \mathrm{Im} \Sigma (\omega)\propto \omega^2$ expected for a Fermi liquid. Ba$_2$RuO$_4$/STO acquires additional kink-like feature in the real part at the energy scale of $15\pm10$ meV near $\mathbf{k}=(0,\pi)$ (red line cut). This flattens the $\gamma$ band and pushes the vHs slightly below the Fermi level. \label{fig:fig3}}
\end{center}
\end{figure}

A schematic of the strain evolution of the $\gamma$ FS is shown in Fig. 2a, where the vertical axis is the effective change in the chemical potential of the $\gamma$-band relative to bulk Sr$_2$RuO$_4$. The change in FS topology cannot be described simply as a rigid shift of the bulk bands; the Fermi surfaces and density of states shown in Fig. 2a and 2b are generated from a generalized tight binding model whose parameters are varied to fit the different strained samples \cite{Supplemental}. The filling of the $\gamma$ band arises from inter-orbital electron transfer from the $d_{xz}$ and $d_{yz}$ orbitals into the $d_{xy}$ orbital; the total number of electrons in all three bands remains constant at 4.00 $\pm$ 0.05 (Fig. 2c). Although density functional calculations indicate that the spin-orbit interaction is non-negligible \cite{Haverkort2008}, we could not directly resolve any spin-orbit split bands, possibly due to impurity scattering and/or experimental resolution. Therefore, we simply utilize a typical 3-band tight-binding model to parameterize our data. The Lifshitz transition has a profound impact on the electron-hole susceptibility, as shown by the Lindhard susceptibility for the 2D $\gamma$ band and the 1D $\alpha$ and $\beta$ bands calculated using a tight binding parameterization of the experimental FS wavevectors and dispersion. Only intraband scattering for $\gamma$ is considered, while both intra- and inter-band scattering between the 1D $\alpha$ and $\beta$ bands is allowed. For the 1D bands, $\chi_{\alpha,\beta}(\mathbf{q}, \omega = 0)$ is relatively independent of strain, except for the reduced nesting in Ba$_2$RuO$_4$ due to its stronger two-dimensionality. For the $\gamma$ band, however, $\chi_{\gamma}(\mathbf{q}, \omega = 0)$ exhibits dramatic changes with strain, where $\chi_{\gamma}(\mathbf{q} = (0,0))$ is strongly enhanced  approaching the Lifshitz transition. There is also a corresponding increase of $\chi_{\gamma}(\mathbf{q} = (\pm\pi,\pm\pi))$, since that wavevector connects the vHs at $(0,\pi)$ and $(\pi,0)$ to symmetry-equivalent pairs. This detailed parameterization of the electronic structure and susceptibility at different strain states should provide valuable input to make falsifiable predictions for the behavior of the superconducting state with strain, as well as help to distinguish which bands are most relevant to superconductivity \cite{Raghu2010, Huo2013, Mazin1997}.

\begin{figure}
\begin{center}
\includegraphics[width=3.375in]{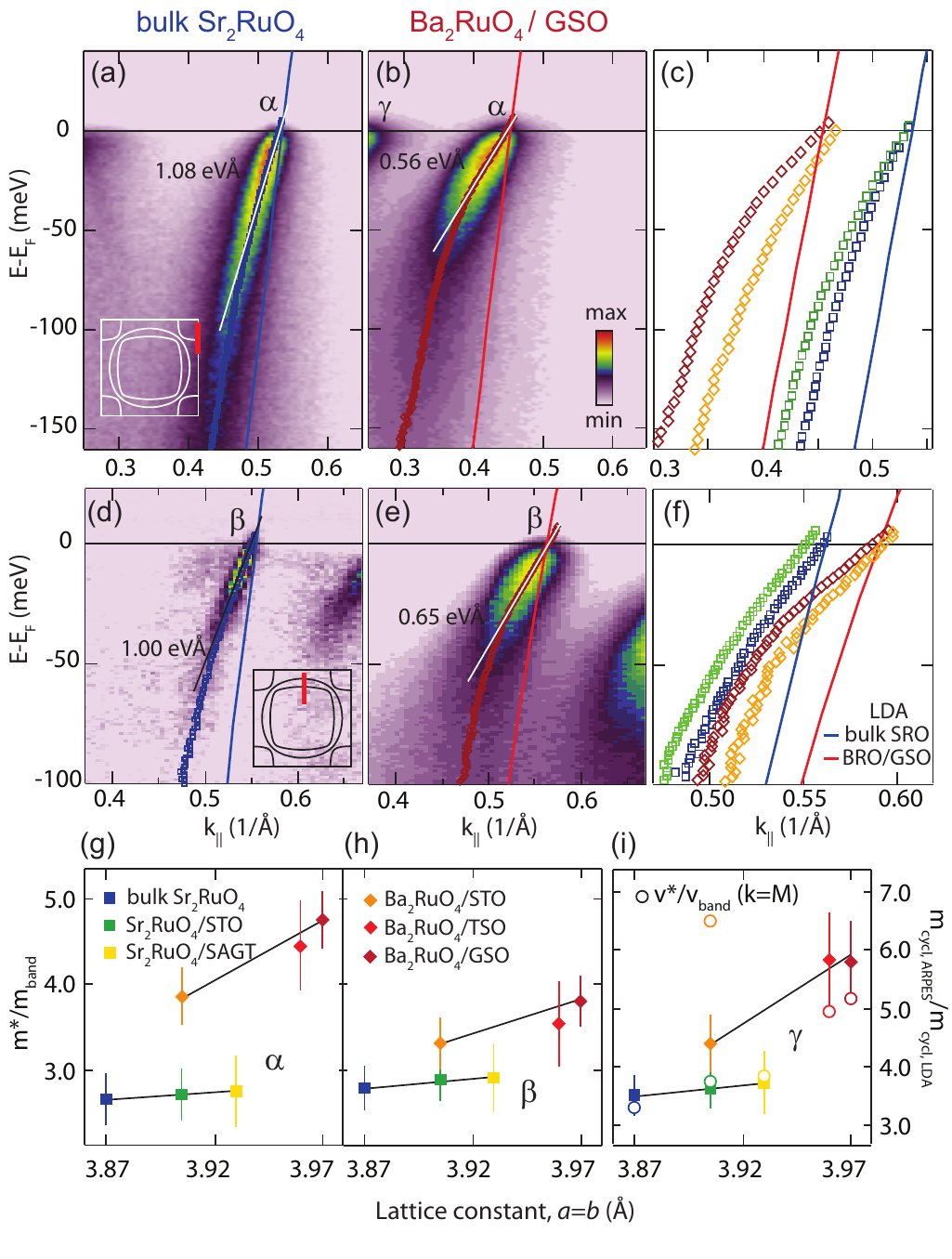}
\caption{
Dispersion $E(k)$ of the $\alpha$ band along the BZ boundary (a-c) and $\beta$ band along (0,0)-(0,$\pi$) (d-f). Spectral weight is shown for the single crystal Sr$_2$RuO$_4$ (a,d) and Ba$_2$RuO$_4$/GSO (b,e). The colors of the open symbols in (c) and (f) are consistent with the colors in (g--i). Quasiparticle renormalizations for the $\alpha$-band (g) and $\beta$-band (h) have strong monotonic dependence on the strain value. QP renormalization for the $\gamma$ band (i), calculated as a ratio of the LDA bandwidth to ARPES bandwidth from the corresponding tight-binding fits, shows a similar monotonic increase as a function of tensile strain. The open circles in (i) show the band renormalization from the slope of the real part of self-energy $1-\partial \mathrm{Re} \Sigma(\omega) / \partial \omega$ near $\mathrm{E_F}$ calculated at $\mathbf{k}=(\pi,0)$. The deviation for Ba$_2$RuO$_4$/STO is due to the band flattening shown in Fig. 3b.
\label{fig:Bands}}
\end{center}
\end{figure}

The impact of the topological transition is clearly evident in the electrical resistivity and in spectroscopic signatures. Between $\mathrm{T_{c}}$ and 25 K, bulk Sr$_2$RuO$_4$ behaves as an ideal Fermi liquid with a $T^2$ resistivity and moderate correlations \cite{Hussey1998, Bergemann2003}. This $T^2$ resistivity is also observed for Sr$_2$RuO$_4$ and Ba$_2$RuO$_4$ films on either side of the Lifshitz transition (Fig. 3a). Close to the Lifshitz transition, however, we observe $\rho(T) \propto T^{1.4 \pm 0.1}$ up to approximately 25 K in Ba$_2$RuO$_4$ / STO as shown in Figs. 3a and 3c, consistent with a previous doping-dependent study \cite{Kikugawa2004}. The quasiparticle dispersion at $(\pi,0)$ also exhibits a deviation from the calculated band structure precisely at the vHs for Ba$_2$RuO$_4$ / STO, as shown in Fig. 3b. At other strain states, the experimental dispersion at $(\pi,0)$ can be well described by a tight binding fit. At the critical strain state, however, the dispersion exhibits an anomalous flattening which deviates strongly from both the LDA calculations and the tight binding parameterization, and cannot be ascribed to the finite experimental resolution \cite{Supplemental}. This can be represented by a deviation of $\Sigma'(w)$ at $\mathbf{k} = (\pi,0)$ from a linear dependence at low energy expected for a conventional Fermi liquid and observed at other locations in momentum space for Ba$_2$RuO$_4$ / STO (Fig. 3e). Since the low-energy electronic structure is highly two-dimensional, the measured quasiparticle properties in Ba$_2$RuO$_4$ / STO appear to be unaffected by any finite thickness effects \cite{Supplemental}. The thin films presented here are non-superconducting, with residual resistivities $\rho_{0} \approx 10^{-5} \ \Omega\cdot$cm, although recent upgrades to the growth chamber should allow us to ultimately achieve superconducting films, as has been reported in unstrained thin films grown on LSAT \cite{Krockenberger2010}.

Given the deviations from canonical Fermi liquid behavior, it is natural to investigate whether the strength of quantum many-body interactions is likewise peaked at the Lifshitz transition. This is shown in Fig. 4, where the measured quasiparticle dispersions $E(k)$ for the $\alpha$ and $\beta$ bands are shown as a function of in-plane lattice constant. Fig. 4g-i summarize the quasiparticle mass renormalization for all three bands crossing $\mathrm{E_F}$. The mass enhancements were calculated using the dispersion along the line cuts for the $\alpha$ and $\beta$ bands and averaged over the full BZ for the $\gamma$ band \cite{Supplemental}. It has been established from both quantum oscillations and prior ARPES measurements that the mass renormalization of $\alpha$ and $\beta$ bands in bulk Sr$_2$RuO$_4$ is approximately 2.5--3 \cite{Bergemann2000, Shen2007}, consistent with our measurements on single crystals of Sr$_2$RuO$_4$. The strength of this renormalization is, however, dramatically enhanced when increasing the Ru-O bond length and substituting the larger A-site cation. Increasing the bond distance by 2.6 \% when going from bulk Sr$_2$RuO$_4$ to Ba$_2$RuO$_4$ on GSO, increases the effective mass of the $\alpha$ band by nearly a factor of 2, far larger than expected than from LDA, which predict less than a 10\% change in $v_{F}$ between these two materials (Fig. 4g), yet a noticeable jump in the renormalization occurs when changing from the Sr to Ba cation at the same lattice constant (SrTiO$_{3}$). In the $\alpha$ and $\beta$ bands, a significant component of the mass enhancement arises from a kink in the dispersion (presumably due to electron-boson coupling) around 80 $\pm$ 40 meV. Nevertheless, even the dispersion at higher binding energies (greater than 100 meV) is substantially renormalized in going from bulk Sr$_{2}$RuO$_{4}$ to Ba$_{2}$RuO$_{4}$ / GSO (a factor of 1.9$\pm$0.2 and 1.8$\pm$0.4 for the $\alpha$ and $\beta$ bands, respectively). It is important to note that the mass enhancement is not peaked at the Lifshitz transition, but rather increases monotonically with Ru-O bond distance, consistent with the increase of correlations from the local repulsion $U/t$ and the Hund's coupling \cite{Mravlje2011,DeMedici2011}.

At this point, we compare strain-induced modifications to prior carrier doping studies \cite{Shen2007}. One advantage of  strain is the potential to investigate its impact on superconductivity and T$_{c}$, whereas chemical disorder destroys superconductivity. Like in doped Sr$_{2-y}$La$_{y}$RuO$_{4}$, we observe signatures of criticality (e.g. $\rho \propto T^{1.4}$ behavior) at low temperatures near the Lifshitz transition \cite{Kikugawa2004}. Some effects of criticality might be partially masked by disorder which can be improved in future generations of thin films. The impact of electron doping on the electronic structure of Sr$_{2-y}$La$_{y}$RuO$_{4}$ could be well described by a simple rigid band shift model, and there was no change in the mass renormalization, even past the Lifshitz transition. In contrast, epitaxial strain impacts the electronic structure in more profound ways, including inducing large increases in the mass renormalization (Fig. 4), and an unexpected band flattening near the Lifshitz transition (Fig. 3b and 3d). The strength of the low-energy kink around 80 meV in the $\alpha$ and $\beta$ bands are also greatly enhanced in Ba$_{2}$RuO$_{4}$ versus Sr$_{2}$RuO$_{4}$, suggesting an increased electron-phonon interaction, which was not reported in Sr$_{2-y}$La$_{y}$RuO$_{4}$. In comparison to the prior work on uniaxial strain \cite{Hicks2014}, our calculations indicate that the impact to the electronic structure along the strained direction is comparable to the effects of biaxial strain. However, under uniaxial strain C$_4$ symmetry is broken, and therefore one pair of van Hove singularities is lowered, while the orthogonal pair is raised in energy. Nevertheless, the uniaxial strain experiments suggest that superconductivity may be strongly intertwined with lowering the vHs, and therefore we speculate that biaxial strain could likewise be a promising route towards enhancing T$_\mathrm{c}$, as will be addressed by a future theoretical study \cite{Hsu2016}.

Our work is the first demonstration of controlling Fermi surface topology and quantum many-body interactions in ruthenates via epitaxial strain engineering, opening the door to future possibilities for engineering quantum many-body ground states in a disorder-free manner to explore enhanced superconductivity, quantum criticality, or electronic nematic states. Tuning the $\gamma$ FS sheet precisely to the Lifshitz transition allows us to place the system at the onset of quantum criticality and observe deviations from canonical Fermi liquid behavior. Our work demonstrates strong inter-orbital electron transfer between the different $t_{2g}$ orbitals with increasing strain, and a topological transition in the $\gamma$ FS sheet. The detailed parameterization of the evolution of the Fermiology and mass renormalization should allow for testable theoretical predictions for changes in the superconducting state with epitaxial strain. 

We thank A. Georges, Y.T. Hsu, and E.A. Kim for helpful discussions and suggestions. This work was primarily supported by Air Force Office of Scientific Research (Grants No. FA9550-12-1-0335 and No. FA2386-12-1-3013). Support from the National Science Foundation was through the Materials Research Science and Engineering Centers program (DMR-1120296, the Cornell Center for Materials Research). This work was performed in part at the Cornell NanoScale Facility, a member of the National Nanotechnology Infrastructure Network, which is supported by the National Science Foundation (Grant No. ECCS-0335765). M.U. acknowledges support from a Postdoctoral Fellowship for Research Abroad (No. 24-162) from the Japanese Society for the Promotion of Science.

B.B. and C.A. contributed equally to this work. 

%

\end{document}